\documentclass[12pt]{article}

\usepackage[margin=1in]{geometry} 
\usepackage{amsmath,amsthm,amssymb}
\usepackage{textcomp}
\usepackage[dvips,final, pdftex]{graphicx}         		
\usepackage[greek, english]{babel}                			
\usepackage[hang,bf]{caption}              			
\usepackage{rotating}                     		 		
\usepackage{subfigure}
\usepackage{units}
\usepackage{fancyhdr}
\usepackage{lscape}
\usepackage{amsfonts}
\usepackage{float}
\usepackage{graphicx}
\usepackage{longtable}
\usepackage{fancybox}
\usepackage{setspace} 
\usepackage{units} 
\usepackage{pdfpages}
\usepackage{lscape} 		
\usepackage{pdflscape}
\usepackage{ulem}
\usepackage{enumerate}
\usepackage{upgreek}

\usepackage{listings}

\begin{document}

\title{Representation of incomplete contact problems by half-planes}

\author{H. Andresen$^{\,\text{a,}}$\thanks{Corresponding author: \textit{Tel}.: +44 1865 273811; \protect \\
\indent \indent \textit{E-mail address}: hendrik.andresen@eng.ox.ac.uk
(H. Andresen).}$\,\,$, D.A. Hills$^{\,\text{a}}$, M.R. Moore$^{\,\text{b}}$\\
{\scriptsize{} {$^{\text{a}}$ Department of Engineering Science,
University of Oxford, Parks Road, OX1 3PJ Oxford, United Kingdom}}\\
{\scriptsize{}{$^{\text{b}}$ Mathematical Institute, University
		of Oxford, Andrew Wiles Building, Radcliffe Observatory Quarter,}}\\
{\scriptsize{} {Woodstock Road, OX2 6GG Oxford, UK}}}
\date{}
\maketitle

\begin{abstract}
Methods for finding the optimal choices of the applied remote loads -- the applied normal force, moment, shear force and remote bulk stresses -- needed to solve frictional contact problems in partial-slip using half-plane theory are derived by using data from contacts analysed by the finite element method. While the normal and shear forces and moment are readily found from equilibrium considerations, in order to determine the bulk stresses we must exploit details of the traction ratio and the direct strain within the contact, both of which are readily extracted from simulations. These contact loads enable the formulation of an equivalent half-plane problem for the contact, which can be used to determine much more precise estimates of the slip-zone sizes than are obtainable from direct use of frictional finite element analysis, as aggregated data from the finite element output is employed, and the half-plane analysis will add precision in terms of satisfaction of the laws of frictional slip and stick.\\

\end{abstract}

\section{Introduction} \label{sec:Sec1}

\hspace{0.4cm} All real physical contact problems are between three dimensional objects of finite size.  Incomplete (convex) contacts occur very widely in mechanical engineering, and we would like to be able to study them in great detail, particularly when the interface is subject to friction.  To do this, a powerful representation is to use a formulation in which each body is idealised as a half-plane, but, as a precursor, we must first  reduce a three-dimensional problem to one which is geometrically two-dimensional, and then replace the finite bodies by half-planes which incorporate small ‘bumps’ on the surface whose profiles match those of the original problem \cite{Barber_2018}.

\begin{figure}[t!]
	\centering
	\includegraphics[scale=0.27, trim= 0 0 0 0, clip]{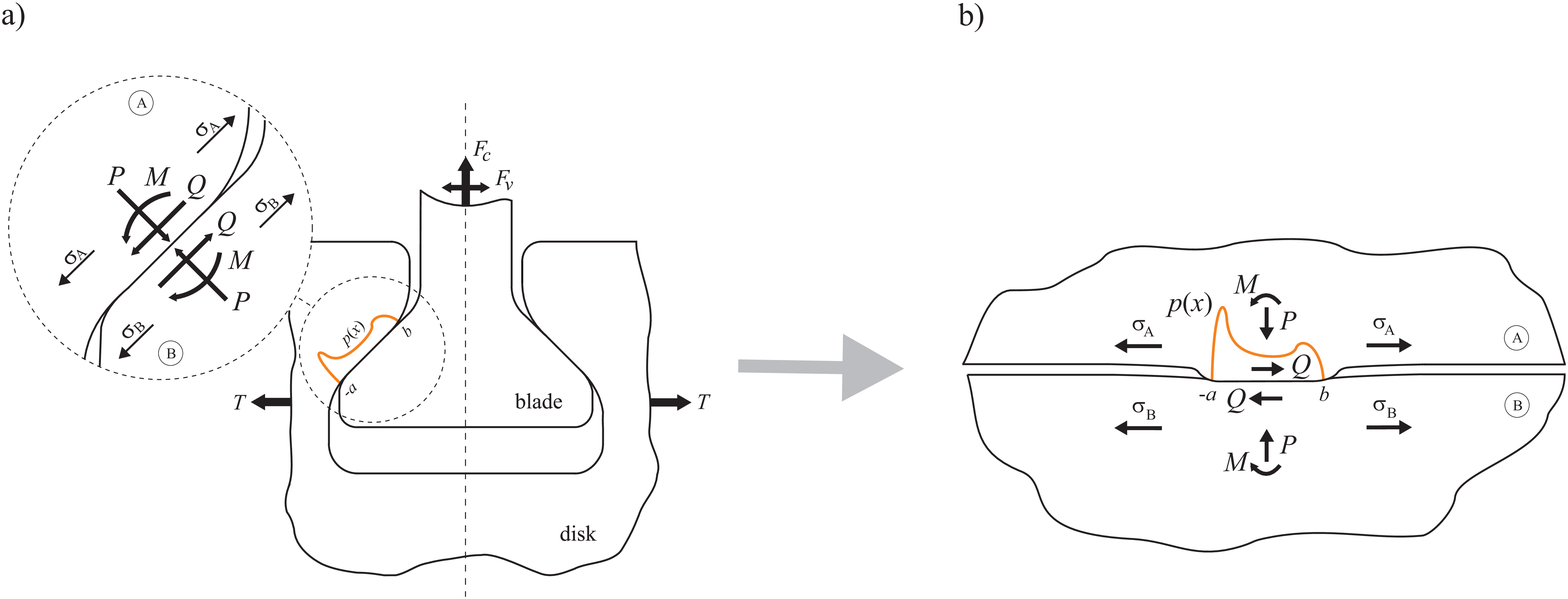}
	\caption{The fan blade dovetail root contact shown in a) is analysed using the half-plane representation shown in b).}
	\label{fig:half_plane_analysis}	
\end{figure}

Figure \ref{fig:half_plane_analysis} a) shows a typical two-dimensional contact problem arising in the fan blade root dovetail of gas turbine engines due to the application of a centrifugal load, $F_C$, vibrational load $F_V$, and expansion load $T$. In its `core' we can identify five quantities which are relevant to the local contact problem, all of which may be functions of time, and four of which control the partial slip problem. 
These are: the normal load, $P$, moment, $M$, shear force, $Q$,
and the differential tension, $\sigma$, developing between bodies A and B, where $\sigma = \sigma_{\text{A}}-\sigma_{\text{B}}$, Figure \ref{fig:half_plane_analysis}.

The normal contact problem is defined by the front face geometry (profile) together with the applied normal force, $P$, and applied moment, $M$. It is trivial to deduce these from the finite element output by evaluating 
\begin{equation}\label{normal_equilibrium}
P=\int_{-a}^{b}p(x)\mathrm{d}x \; \text{,}
\end{equation}
\begin{equation}\label{roational_equilibrium}
M=\int_{-a}^{b}p(x)x\mathrm{d}x \; \text{,}
\end{equation}
where $p(x)$ is the contact pressure. Shear tractions, which tend to induce slip, are actually excited by two forms of loading. One is by the application of an external shear force, $Q$, which again is straightforward to deduce as a traction resultant by evaluating 
\begin{equation}\label{tangentialeq}
Q=\int_{-a}^{b}q(x)\mathrm{d}x \; \text{,}
\end{equation}
where $q(x)$ is the shear traction within the contact.

However, there is also a second form of excitation of shear tractions which is clear in the half-plane model, Figure  \ref{fig:half_plane_analysis}  b), namely the effect of the bulk tensions exerted remotely parallel with the surface which, if unequal, produce an (approximately) antisymmetric interfacial shear traction distribution in the half-plane contact. These remote forms of loading are very easy to visualise in the half-plane problem, but when we consider finite problems such as that shown in Figure \ref{fig:half_plane_analysis} a), and where the geometry of individual components looks very different from a half-plane, their ‘equivalent’ is very much harder to interpret. It is the primary function of this paper to show how we may best interpret the output of the finite element analysis of a problem such as that shown in Figure \ref{fig:half_plane_analysis}  a) to enable finding the optimal choices for the remote stresses applied to the half-plane contact problem shown in Figure \ref{fig:half_plane_analysis} b).

In a structural engineering sense, the prototypical problem shown in Figure \ref{fig:half_plane_analysis}  a) is redundant, because the two contact flanks mean that the loads carried on each individual contact cannot be determined by considerations of equilibrium alone, and we would therefore normally begin our analysis by forming a finite element model of the problem using a commercial code. This code outputs the local contact pressure, $p(x)$, and shear tractions, $q(x)$. These outputs, along with the geometry of the structure together with the coefficient of friction, provide a connection between the external loads and the local response acting on the contact interface

\begin{equation}\label{matrix}
\left\{ \begin{array}{c}
P\\
Q\\
M\\
\sigma_{\text{A}}\\
\sigma_{\text{B}}
\end{array}\right\} =\left[\begin{array}{ccc}
e_{11}  \quad e_{12} \quad e_{13}
\\  e_{21}  \quad e_{22} \quad e_{23}
\\  e_{31}  \quad e_{32} \quad e_{33}
\\  e_{41}  \quad e_{42} \quad e_{43}
\\  e_{51}  \quad e_{52} \quad e_{53}
\end{array}\right]\left\{ \begin{array}{c}
F_{C}\\
F_{V}\\
T
\end{array}\right\} \; \text{.}
\end{equation}

Since the loads $P$, $Q$, $M$, $\sigma_{A}$ and $\sigma_{B}$ are the necessary inputs to any half-plane analysis (cf. Figure \ref{fig:half_plane_analysis} b)), it is the intention of this paper is to provide means of determining this connection, here indicated by the matrix entries $e_{ij}$, for $i=1,2,3,4,5$ and $j=1,2,3$. While the first nine matrix entries ($e_{1j}$, $e_{2j}$, $e_{3j}$) manifest themselves in evaluating Eqs. (\ref{normal_equilibrium}, \ref{roational_equilibrium}, \ref{tangentialeq}), determining the remaining entries ($e_{4j}$, $e_{5j}$) is far more complicated. But good estimates of $\sigma_{\text{A}}$ and $\sigma_{\text{B}}$ are needed in order to construct, firstly, an analytical model of the individual contact using a model based on half-plane domains, where the shape of the contacting surfaces is the same as the corresponding function for the prototypical contact. The half-plane analysis becomes advantageous because the computationally justifiable resolution of the finite element model is too poor to investigate thoroughly the partial slip problem in full detail \cite{Banerjee_2009}. Secondly, by knowing the individual remote stresses present in each body, a far more accurate representation of the propagation stresses for potentially initiated cracks in the prototype can be applied when a simpler laboratory set up is used in an experimental investigation.

Once the values for all five contact loads, $P$, $Q$, $M$, $\sigma_{\text{A}}$ and $\sigma_{\text{B}}$ are found, the whole of the family of solutions for partial-slip analysis becomes available,
for example the pioneering solutions of Cattaneo and Mindlin \cite{Cattaneo_1938}, \cite{Mindlin_1949}
and more recent developments \cite{Andresen_2019_3}, including dislocation-based formulations \cite{Moore_2018}, \cite{Moore_2018_2}. We wish to emphasise that, for reasons which will become clear, the finite problem is investigated under fully adhered conditions. This simply allows us to exploit specific properties of the tractions arising during loading which are relevant in identifying the tension components.  However, this does not imply that any subsequent analysis must be carried out with this assumption. We will develop the problem in stages. 

\section{Proportional loading: no moment developed}\label{nomoment}
\begin{figure}[b!]
	\centering
	\includegraphics[scale=0.4, trim= 0 0 0 0, clip]{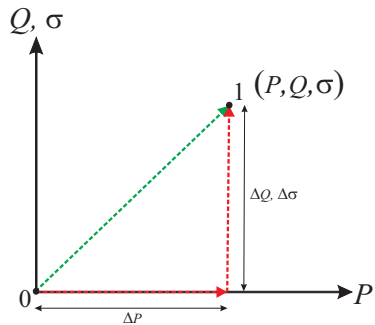}
	\caption{Sequential loading (red) and proportional loading (green) for a $P,Q,\sigma$-problem.}
	\label{fig:PQSimga_Space}	
\end{figure}

\hspace{0.4cm}
 Figure \ref{fig:PQSimga_Space} shows two very common loading sequences for a $P, Q, \sigma$-problem neglecting the effect of a moment. In red we see a sequential order of load application and in green we see proportional loading from one load point to another. The sequence in which the loads are applied greatly affects the procedure we wish to develop. In real problems the external loads may not be applied separately, in the sense of forming the contact
 first, and then applying the forces which excite shear but, since it is of interest academically, we discuss details of sequential loading in detail in Appendix \ref{S-sequentialloading}. 
 
 Usually, the external loads applied have the effect of exerting all local loads proportionally, so that 
\begin{equation}\label{proploading}
\frac{\mathrm{d}P}{P}=\frac{\mathrm{d}M}{M}=\frac{\mathrm{d}Q}{Q}=\frac{\mathrm{d}\sigma_{\text{A}}}{\sigma_{\text{A}}}=\frac{\mathrm{d}\sigma_{\text{B}}}{\sigma_{\text{B}}}\; \text{.}
\end{equation}

The problem becomes significantly more complicated when there is a moment present, and it
is not possible to treat the problem as comprehensively (see \textsection\ref{moment})
so here we will begin by looking in some detail at problems where the
moment may be neglected. Consider the problem shown in Figure \ref{fig:half_plane_analysis} a) and b), and suppose
that the contacting bodies are elastically similar. In our analysis we will neglect the shear tractions induced by the application of the normal load as they will be much smaller in magnitude than the contact pressure when the bodies are elastically similar and arise only because the contacting bodies usually have different domain shapes. While the remarks about deducing the values of
normal load and shear force ($P$, $Q$) as traction resultants in \textsection \ref{sec:Sec1} still apply, we now need a means of finding the individual remote tensions ($\sigma_{\text{A}}, \sigma_{\text{B}}$).
To simplify even further, we will look at the reduced case of a contact in which the indenter is also symmetric, so that in the half-plane formulation of figure \ref{fig:half_plane_analysis} b), $b = a$. Moreover, this also implies that we neglect the presence of the vibrational external load, $F_{V}$, which somewhat simplifies the matrix system Eq. (\ref{matrix}).

We shall begin by describing a methodology for finding the bulk tension difference, $\sigma = \sigma_{A}-\sigma_{B}$. Suppose that we have a contact whose current
half-width is $a$, and we make a small change, $\Delta P$, in normal
load. The corresponding change in contact pressure is given by \cite{Hills_2011}
\begin{equation}\label{ec_05}
\Delta p(x)=\frac{\Delta P}{\pi\sqrt{a^{2}-x^{2}}}\; \text{.}
\end{equation}

If at the same time there are small changes in shear force, $\Delta Q,$
and differential bulk tension, $\Delta\sigma$, the change in shear traction, $\Delta q(x),$ generated is given
by \cite{Hills_2011}
\begin{equation}\label{ec_02}
\Delta q(x)=\frac{\Delta Q}{\pi\sqrt{a^{2}-x^{2}}}+\frac{\Delta\sigma x}{4\sqrt{a^{2}-x^{2}}}\; \text{.}
\end{equation}

If full stick is to be maintained during a positive load increment we require that \cite{Hills_2011}
\begin{equation}\label{full_stick_inequality}
\frac{|\Delta Q|}{\Delta P}+\frac{\pi}{4}\frac{a \Delta\sigma}{\Delta P}<f\; \text{.}
\end{equation}

We wish to find the change in interfacial shear traction moving linearly
from an unloaded state $0$ to state $1$ {[}$P_{1},Q_{1},\sigma_{1}${]}\footnote{For simplicity, we will omit the subscript for the load state from here on.}, see Figure \ref{fig:PQSimga_Space}, under conditions which ensure full stick, i.e. inequality \eqref{full_stick_inequality} is satisfied. Suppose we also know, from the contact geometry, the contact law, $a=g(P)$, and we denote the contact pressure distribution
at normal load $P$ by $p(x,P)$. The contact pressure at state 1 is given by integrating Eq. (\ref{ec_05}),
\begin{equation}\label{eqn:PsymmIntegral}
p(x)=\frac{1}{\pi}\int_{P_{x}}^{P}\frac{\mathrm{d}\tilde{P}}{\sqrt{a^{2}-x^{2}}} \; \text{,}
\end{equation}
where $g(P_{x})=x$. We make the change of variable $\tilde{a} = g(\tilde{P})$ in the integrand, which gives
\begin{equation}\label{pressureintegral}
p(x)=\frac{1}{\pi}\int_{x}^{a}\frac{1}{g^{'}(\tilde{P})}\frac{\mathrm{d}\tilde{a}}{\sqrt{\tilde{a}^{2}-x^{2}}} \; \text{,}
\end{equation}
where a prime indicates differentiation with respect to argument.

Now we consider the developing shear tractions and study first, for
simplicity, the effect of a simultaneously exerted shear force alone.
Then, from Eq. (\ref{full_stick_inequality}), we require the change in normal and shear load to be
\begin{equation}
\frac{\Delta Q}{\Delta P}=\frac{Q}{P} = \lambda < f\; \text{,}
\end{equation}
where the $\Delta$ terms indicate the change in load from state $0$ and $1$. Hence 
\begin{equation}\label{nobulkttracs}
\Delta q(x)=\frac{\lambda\Delta P}{\pi\sqrt{a^{2}-x^{2}}}\; \text{,}
\end{equation}
so that, if we applied a shear force only, 
\begin{equation}\label{lambda}
q(x, P)=\lambda p(x,P)\; \text{.}
\end{equation}

If we introduce a differential bulk tension as well, so that both develop concurrently (satisfying the `proportional loading' property), and set
\begin{equation}\label{eta}
\eta=\frac{a\Delta\sigma}{\Delta P}= \frac{a \sigma}{P}\; \text{,}
\end{equation}
($\Delta\sigma\equiv\sigma$ and $\Delta P\equiv P$ as we start from an unloaded state),
 we see that Eq. (\ref{nobulkttracs}) is now replaced by
\begin{equation}\label{bulkttracs}
\Delta q(x)=\frac{\lambda\Delta P}{\pi\sqrt{a^{2}-x^{2}}}+\frac{x\eta\Delta P}{4 a \sqrt{a^{2}-x^{2}}}\; \text{.}
\end{equation}

Thus after integrating Eq. \ref{bulkttracs} with respect to $P$, we find that 
\begin{equation}
q(x, P)=\int_{P_{x}}^{P}\left[\frac{\lambda}{\pi\sqrt{a^{2}-x^{2}}}+\frac{x\eta}{4a\sqrt{a^{2}-x^{2}}}\right]\mathrm{d}\tilde{P}=\left[\frac{\lambda}{\pi}+\frac{x\eta}{4a}\right]\int_{x}^{a}\frac{\mathrm{d}\tilde{a}}{g^{'}(\tilde{P})\sqrt{\tilde{a}^{2}-x^{2}}}\; \text{,}
\end{equation}
and, by comparison with Eq. \eqref{pressureintegral}, we find the traction ratio as
\begin{equation}\label{tractionratio}
\frac{q(x, P)}{p(x,P)}=\lambda+\frac{\eta \pi}{4a} x\; \text{,}
\end{equation}
which clearly reduces to Eq. \eqref{lambda} in the case that the bulk tension difference is negligible.

As expected, the shear traction is related to a scaled form of the
contact pressure. Thus, given output data for $p(x)$ and $q(x)$ from a finite element simulation or otherwise, we are able to find the implied values of $\Delta\sigma$ and $\Delta Q$, or $\sigma$ and $Q$ by considering the traction ratio across the contact interface, Eq. \eqref{tractionratio}. We shall illustrate this in detail in the upcoming examples.

We now turn our attention to finding the \textit{sum} of the bulk stresses, $\sigma_{\text{A}}+\sigma_{\text{B}}$,
and start by writing down the total surface strains arising along the contact interface in the presence of a normal load, $P$, a shear force, $Q$, differential tension, $\sigma = \sigma_{\text{A}}-\sigma_{\text{B}}$, and the \textit{average} remote bulk loads, $\nicefrac{\sigma_{\text{A}} + \sigma_{\text{B}}}{2}$,
\begin{equation}\label{total_directstrain_propL}
\varepsilon_{xx}(x)= \varepsilon_{xx}^{P}(x) + \varepsilon_{xx}^{Q}(x) + \varepsilon_{xx}^{\sigma} + \varepsilon_{xx}^{\sigma_{\text{A}}+\sigma_{\text{B}}}(x) \;, \;-a<x<a\text{.}
\end{equation}

The contribution to the total direct strain from a normal force, $P$, can be written as
\begin{equation}\label{PMstrain}
\varepsilon_{xx}^{P}(x)=\frac{\left(1-2\nu\right)}{(1-\nu)\,E^{*}}\,p(x)\;, -a<x<a\;\text{,}
\end{equation}
where $E^{*}=E/\left(1-\nu^{2}\right)$, $E$ is Young's modulus
and $\nu$ is Poisson's ratio.
As we have found a solution under no-slip conditions, we can also
find the implied surface strains resulting from the effects of an applied shear force, $Q$, and the shear tractions arising from differential bulk tension, $\sigma$, which gives 
\begin{equation}\label{Qstrain}
\varepsilon_{xx}^{Q}(x)=-\frac{2}{\pi E^{*}} \, \lambda \, \int_{-a}^{a}\frac{\displaystyle{\, p\left(\xi\right)}}{x-\xi} \mathrm{d}\xi\;, -a<x<a\; \text{,}
\end{equation}
and
\begin{equation}\label{sigmastrain}
\varepsilon_{xx}^{\sigma}(x)=-\frac{2}{\pi E^{*}}\,  \frac{\eta\pi}{4a} \, \int_{-a}^{a}\frac{\displaystyle{p\left(\xi\right) \,\xi}}{x-\xi} \,\mathrm{d}\xi\;,-a<x<a\; \text{.}
\end{equation}

The total strain distribution within the fully stuck contact will
therefore have four contributions, \textit{viz}. the three just derived
$\varepsilon_{xx}^{P}(x),\varepsilon_{xx}^{Q}(x),\varepsilon_{xx}^{\sigma}(x)$
together with the effect of the \textit{average} remote bulk loads, i.e.
\begin{equation}\label{sigmasumstrain}
\varepsilon_{xx}^{\sigma_{\text{A}}+\sigma_{\text{B}}}\left(x\right)=\frac{\sigma_{\text{A}}+\sigma_{\text{B}}}{2E^{*}}\; \text{.}
\end{equation}

Hence, given data for the contact pressure, $p(x)$, from finite element results or otherwise, we can now calculate the individual strain contributions given by Eqs. (\ref{PMstrain}, \ref{Qstrain}, \ref{sigmastrain}). A comment is appropriate in relation to the evaluation of the integral. We will require the
surface strains within the contact ($-a\leq x\leq a$) so that the
integral is interpreted in the Cauchy principal value sense. It will therefore be appropriate
to use a numerical integration scheme to evaluate Eqs. (\ref{Qstrain}, \ref{sigmastrain}), such as Gauss-Chebyshev \cite{Hills_1996}.
In order to implement this the integrand should be evaluated at the
prescribed integration points. Therefore either the mesh used
in the finite element analysis should be chosen so that the nodes
lie at the desired coordinates or an appropriate interpolation
scheme might be used, such as a variant of that developed by Krenk \cite{Krenk_1975}. As we have seen, the total direct surface strain is a combination of the strains arising from the tractions and those emanating directly from the sum of the bulk stresses, and so the latter may be found as
\begin{equation}\label{SigmaSigma}
 \sigma_{\text{A}}+\sigma_{\text{B}} = 2 E^* \left[ \varepsilon_{xx}(x) -\varepsilon_{xx}^{P}(x) - \varepsilon_{xx}^{Q}(x) - \varepsilon_{xx}^{\sigma} \right]\text{.}
\end{equation}

In summary, the following steps are needed in order to find the values of $P$, $Q$, $\sigma_{\text{A}}$ and $\sigma_{\text{B}}$ under proportional loading of a symmetric indenter in the absence of a moment:
\begin{enumerate}[(i)]
	\item Find values for the contact tractions $p(x)$ and $q(x)$ from simulations or otherwise. 
	\item The contact resultants, $P$ and $Q$, are found from integrating the tractions using Eqs. (\ref{normal_equilibrium}, \ref{tangentialeq}).
	\item The gradient of the linear function $\nicefrac{q(x,P)}{p(x,P)} = \lambda+\frac{\eta \pi}{4a} x$ reveals the value of $\sigma = \sigma_{\text{A}} - \sigma_{\text{B}}$, using Eq. \eqref{eta}.
	\item Evaluate the individual contributions to the direct strain, $\varepsilon_{xx}$, using Eqs. (\ref{PMstrain}, \ref{Qstrain}, \ref{sigmastrain}) and subtract those from the numerical output for the total strain to determine the sum of the bulk stresses using Eq. \eqref{SigmaSigma}.
	\item Finally, output $\sigma_{\text{A}} = \nicefrac{\sigma}{2} + \nicefrac{\sigma_{\text{A}} + \sigma_{\text{B}}}{2}$ and $\sigma_{\text{B}} = \nicefrac{\sigma_{\text{A}} + \sigma_{\text{B}}}{2}- \nicefrac{\sigma}{2}$.
\end{enumerate}

We thus have all the required information in the matrix system Eq. (\ref{matrix}): given inputs $F_{C}$ and $T$, we output $p(x)$ and $q(x)$ and are then able to follow (i)-(v) to recover the loads necessary for an equivalent half-plane analysis.

One of the advantages of considering the simplified case in which we neglect a moment and any geometrical asymmetry is that it is often possible to find closed-form solutions for the equivalent half-plane model, so we are able to illustrate this methodology for cases in which we already know the relationships in Eq. (\ref{matrix}), which we shall pursue in \textsection \ref{Hertzexample}. We shall then conclude our discussion of the symmetric problem by discussing an alternative to step (iii) in the methodology that will be of great value to our extension to problems involving a moment or asymmetry in \textsection \ref{moment}.

\subsection{Demonstration of the methodology for a Hertzian contact}
\label{Hertzexample}

\begin{figure}[b!]
	\centering
	\includegraphics[scale=0.34, trim= 0 0 0 0, clip]{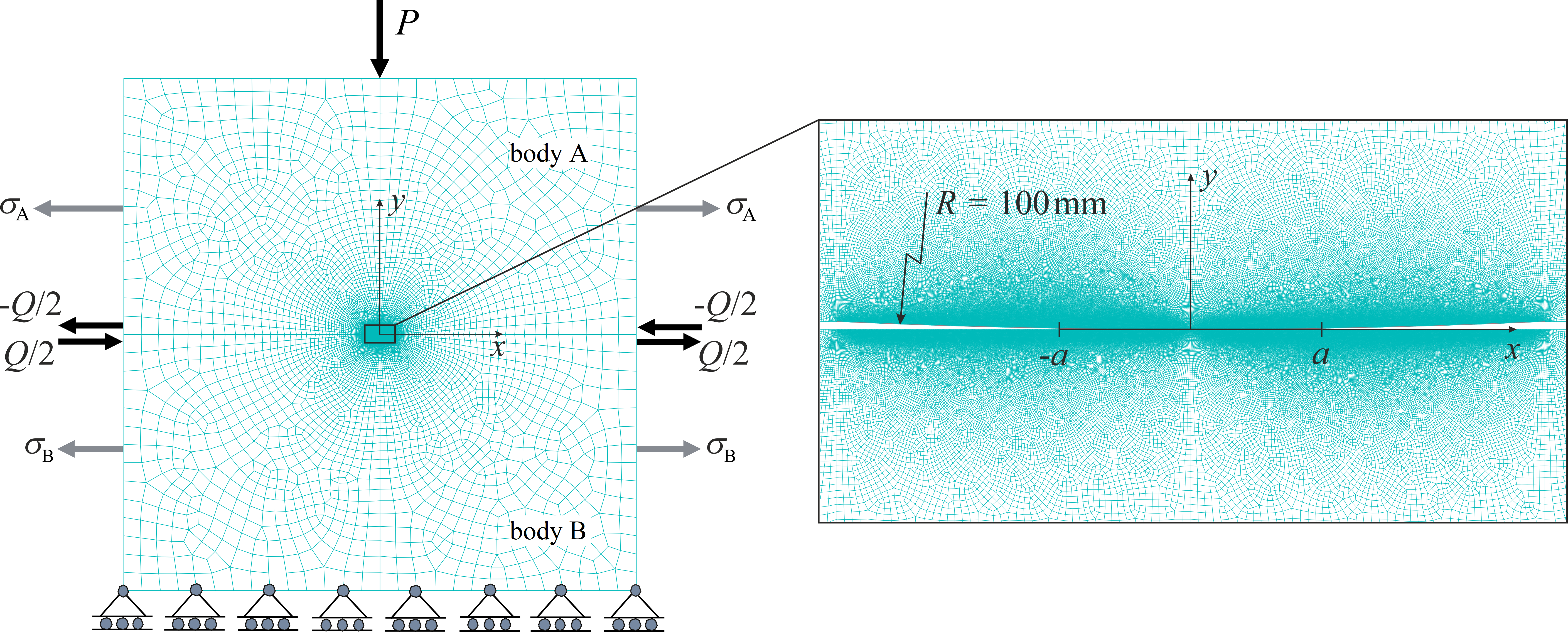}
	\caption{FE-Model of two half-planes connected by a Hertzian-type geometry subject to loads $P$, $Q$, $\sigma_{\text{A}}$, and $\sigma_{\text{B}}$.}
	\label{fig:Hertz_Fe_model}	
\end{figure}

\hspace{0.4cm} The motivation behind developing this method is to be able to find the remote stresses applied, which in many real problems such as that shown in Figure \ref{fig:half_plane_analysis}  a) are difficult to interpret. However, in order to illustrate the methodology described by the steps (i)-(v), we shall, as a first example, forgo the wider geometry that leads to the contact as illustrated in Figure \ref{fig:half_plane_analysis} a) and consider the well-known example of a Hertzian contact of two elastically-similar cylinders. We perform finite element simulations of the configuration illustrated in Figure \ref{fig:Hertz_Fe_model} to output the contact tractions $p(x)$ and $q(x)$. Obviously, this is a contrived problem where we actually know the desired values $P$, $Q$, $\sigma_{A}$ and $\sigma_{B}$, but we will recover the remote stresses from the finite element output for the tractions, $p(x)$ and $q(x)$, and the direct surface strain, $\varepsilon_{xx}$, alone, \textit{as if $\sigma_{\text{A}}$ and $\sigma_{\text{B}}$ were unknowns}, i.e. we shall follow steps (iii)-(v) algorithmically to recover the correct values of the bulk stresses. Generally, the finite element output alone, together with an appropriate numerical scheme, suffices to solve the problem, and the closed-form results, if tractable, serve as a point of comparison and simplifies the calculation. 

Figure \ref{fig:Hertz_Fe_model} shows the input configuration for our finite element simulations. We consider two half-planes connected by a Hertzian-type geometry with radius, $R$, over a contact patch spanning $[-a, \quad a]$. Body B is constrained in the vertical direction such that the normal load is supported and shear loading in the lateral direction is balanced between the two bodies.  We thereby acknowledge that the half-planes are only approximated as each body is finite in size, where the proportion of contact half-width to width or height of the two bodies is less than $1/100$. The four loads, $P$, $Q$, $\sigma_{\text{A}}$, and $\sigma_{\text{B}}$ are applied in proportion from an unloaded state under a fully adhered conditions until they reach the desired load state, as shown in Figure \ref{fig:PQSimga_Space}. 

We use the commercial finite element solver Abaqus to solve the contact problem with quadratic quadrilateral and triangular elements under plane strain conditions with a mesh refinement toward the contact interface so that the element size to contact size ratio is less than $1/1000$. The contact interface is meshed such that nodes are coincidental. The contact is modelled as a surface-to-surface interaction with default properties for normal and tangential behaviour suppressing relative lateral motion of the two bodies. 

\begin{figure}[t!]
	\centering
	\includegraphics[scale=0.55, trim= 0 0 0 0, clip]{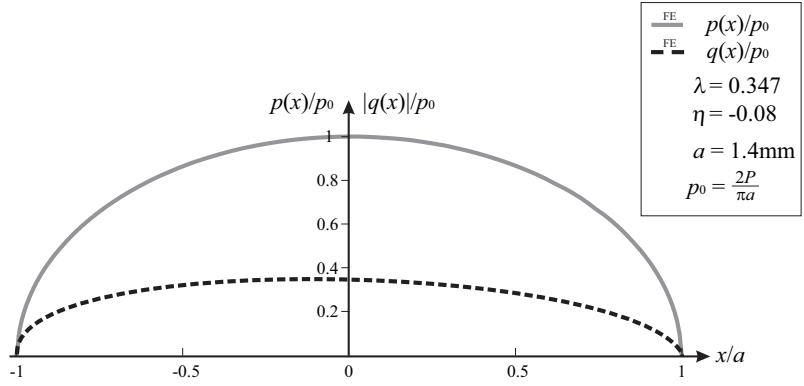}
	\caption{Normal and shear tractions for a Hertzian geometry under proportional loading in fully adhered conditions.}
	\label{fig:Normal_and_shear_traction}	
\end{figure}

The contact tractions extracted from the finite element simulation are shown in Figure \ref{fig:Normal_and_shear_traction}, where the input values of $P$, $Q$ are such that $\lambda = 0.347$. The input for the remote tensions is and $\sigma_{\text{A}} = 80 \nicefrac{\unit{N}}{\unit{mm^2}}$ and $\sigma_{\text{B}} = 180 \nicefrac{\unit{N}}{\unit{mm^2}}$ giving $\eta =-0.08$. Thus, our aim is now to reproduce these by following step (iii).

Given $P$, the contact law $a = g(P)$ for a Hertzian geometry is given by \cite{Hertz_1881}
\begin{equation}
 a = \left(\frac{8 P R}{\pi E^{*}}\right)^{\frac{1}{2}}.
\end{equation}

\begin{figure}[b!]
	\centering
	\includegraphics[scale=0.55, trim= 0 0 0 0, clip]{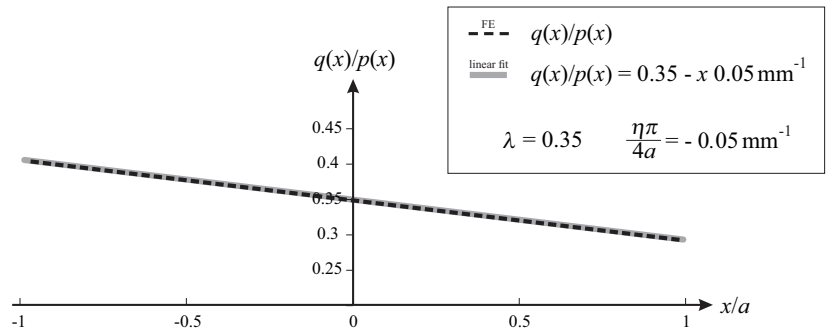}
	\caption{Traction ratio between shear and normal tractions for a Hertzian geometry under proportional loading in fully adhered conditions.}
	\label{fig:tractionratio}	
\end{figure}

Thus, for the input value of $P$ and with $R = 100 \unit{mm}, E^{*} = 225 \; 10^3 \nicefrac{\unit{N}}{\unit{mm^2}}$, the contact size at state $1$ as defined in Figure \ref{fig:PQSimga_Space} is $a = 1.4 \unit{mm}$. We now plot the ratio $q(x)/p(x)$ in Figure \ref{fig:tractionratio} and find a linear fit, which referring back to (iii) finds that
\begin{equation}
 \lambda = 0.35 \quad\mbox{and} \quad \frac{\eta\pi}{4a} = -0.05\unit{mm^{-1}} \Rightarrow \eta = -0.09 \Rightarrow \sigma \approx -111\nicefrac{\unit{N}}{\unit{mm^2}}.
\end{equation}
%\Rightarrow \sigma=-86\nicefrac{\unit{N}}{\unit{mm^2}}

As is evident, we are reliably extracting plausible values of $\lambda$ and $\eta$ from step (iii).

To recover the sum of the bulk stresses, we follow step (iv) and consider the direct strain in the contact interface. We can output the total strain $\varepsilon_{xx}(x)$ directly from the finite element simulations. To evaluate $\sigma_{A}+\sigma_{B}$, we also need the individual contributions $\varepsilon_{xx}^{P}(x), \varepsilon_{xx}^{Q}(x), \varepsilon_{xx}^{\sigma}(x)$. While these can be calculated numerically from the given data for $p(x)$, we note that, for this simple example, we can in fact evaluate them in closed form. By Eq. \eqref{pressureintegral}, the pressure distribution is given by
\begin{equation}
p(x)=\frac{E^{*}}{4 R}\int_{x}^{a}\frac{\tilde{a}\,\mathrm{d}\tilde{a}}{\sqrt{\tilde{a}^{2}-x^{2}}}=\frac{E^{*}}{4R}\sqrt{a^{2}-x^{2}}\; \text{,}
\end{equation}
so that directly evaluating Eq. (\ref{PMstrain}) yields
\begin{equation}\label{directstrain_propL_Hertz}
\varepsilon_{xx}^{P}(x)=\frac{\left(1-2\nu\right)}{(1-\nu)\,E^{*}}\,p(x)=\frac{\left(1-2\nu\right)}{4R\,(1-\nu)}\,\sqrt{a^{2}-x^{2}}\; \text{,}
\end{equation}
while Eq. (\ref{Qstrain}) gives
\begin{equation}\label{directstrain_propL_Hertz_evaluatedQ}
\varepsilon_{xx}^{Q}(x)=-\frac{2}{\pi E^{*}} \, \lambda \, \int_{-a}^{a}\frac{\displaystyle{\, p\left(\xi\right)}}{x-\xi} \mathrm{d}\xi=-\frac{1}{2 R} \frac{Q}{P}\, x\; \text{,}
\end{equation}
and Eq. (\ref{sigmastrain}) gives
\begin{equation}\label{directstrain_propL_Hertz_evaluatedSigma}
\varepsilon_{xx}^{\sigma}(x)=-\frac{2}{\pi E^{*}}\,  \frac{\eta\pi}{4a} \, \int_{-a}^{a}\frac{\displaystyle{p\left(\xi\right)}}{x-\xi} \,\xi\,\mathrm{d}\xi=-\frac{1}{16 R} \frac{\sigma}{P} \left(a^2 - 2 x^2\right)\; \text{.}
\end{equation}

\begin{figure}[b!]
	\centering
	\includegraphics[scale=0.55, trim= 0 0 0 0, clip]{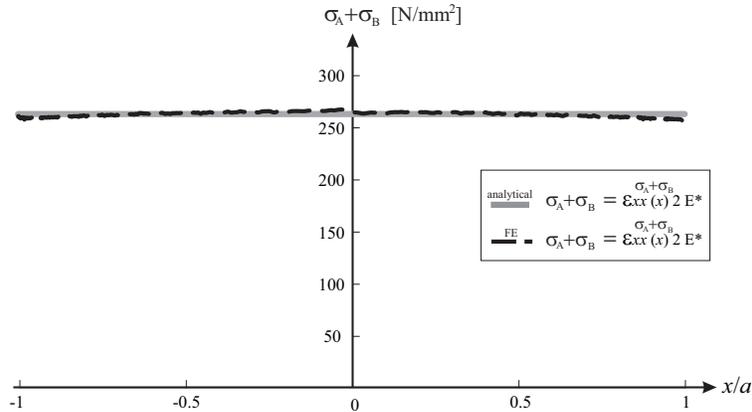}
	\caption{Sum of the bulk stresses, $\sigma_{\text{A}}+\sigma_{\text{B}}$, obtained using Eq. \eqref{directstrain_ana} and finite element output for the direct strain, $\varepsilon_{xx}(x)$.}
	\label{fig:Exx_Distribution_Hertz}	
\end{figure}

Thus, we find that
\begin{equation}\label{directstrain_ana}
\sigma_{\text{A}}+\sigma_{\text{B}} = 2E^{*}\left(\varepsilon_{xx}(x)  + \frac{1}{2 R} \frac{Q}{P}\, x+\frac{1}{16 R} \frac{\sigma}{P} \left(a^2 - 2 x^2\right)-\frac{\left(1-2\nu\right)}{4R\,(1-\nu)}\,\sqrt{a^{2}-x^{2}}\right)\;\text{,}
\end{equation}
where $|x|<a$. Figure \ref{fig:Exx_Distribution_Hertz} illustrates that Eq. \eqref{directstrain_ana} yields a constant value using the analytical description and an almost constant value of $\sigma_{\text{A}}+\sigma_{\text{B}} \approx 260 \nicefrac{\unit{N}}{\unit{mm^2}}$ using the finite element output for the direct strain. The slight curvature in the numerical result might stem from the half-plane approximation in the finite element model, see Figure \ref{fig:Hertz_Fe_model}. 

Therefore, combining our values for $\eta$ and $\sigma_{\text{A}}+\sigma_{\text{B}}$, we finally back out the values
\begin{equation}
 \sigma_{\text{A}} \approx 74 \nicefrac{\unit{N}}{\unit{mm^2}} \; \mbox{and} \; \sigma_{\text{B}} \approx 186 \nicefrac{\unit{N}}{\unit{mm^2}}
\end{equation}
from step (v), which again shows good agreement with the (for this example known) inputs.

With this example, we demonstrated that it is possible to recover all external load inputs,  $P$, $Q$, $\sigma_{\text{A}}$, and $\sigma_{\text{B}}$, from the tractions and direct strains within the contact interface alone. In \textsection \ref{moment}, we will show how this methodology may be adapted to find answers to the same questions in more complicated problems, for instance when a moment is present or for redundant systems such as the dovetail geometry for gas turbine fan blade joints. 

\subsection{Using asymptotes in step (iii)}
\label{asymptotes}

\hspace{0.4cm} In most real problems, we will not know the contact geometry in a simple
form, but the steps (i)-(v) can still be employed. We conclude by reconsidering step (iii). Given numerical results for $p(x)$ and $q(x)$, there are basically two routes which may be followed. The most obvious method is to plot the ratio $\nicefrac{q(x)}{p(x)}$ across the length of the contact from the finite element results directly, as we did in \textsection \ref{Hertzexample}. This ratio should be linear in $x$ and have an intercept giving the value of $\lambda$ and the gradient $\nicefrac{\eta\pi}{4a}$.
This is the preferred method as it applies averaging of the given numerical results. 

An alternative method, whose extension to the asymmetric case will be exploited in the subsequent section, is to look in detail at the traction-ratio towards the ends of the contact, where the contact edge values are given by 
\begin{equation}\label{trLHS}
\frac{q(a^{-})}{p(a^{-})}=\lambda+\frac{\eta\pi}{4}=\frac{Q}{P}+\frac{\pi a\sigma}{4 P}\; \text{,}
\end{equation}
and 
\begin{equation}\label{trRHS}
\frac{q(-a^{+})}{p(-a^{+})}=\lambda-\frac{\eta\pi}{4}=\frac{Q}{P}-\frac{\pi a\sigma}{4 P}\; \text{.}
\end{equation}

If we move the origin to the contact edge and set $s=a+x$, $t=a-x$, we can express the normal and shear tractions near the contact edges in asymptotic form as \cite{Fleury_2017}
\begin{equation}
p(s)\approx K_{P}^{-}\sqrt{s} \; \text{,} \qquad p(t)\approx K_{P}^{+}\sqrt{t}\; \text{,}
\end{equation}
\begin{equation}
q(s)\approx K_{Q}^{-}\sqrt{s}\; \text{,} \qquad q(t)\approx K_{Q}^{+}\sqrt{t} \; \text{,}
\end{equation}
and as $s,t \rightarrow 0$, the ratio is given as
\begin{equation}
\frac{q(-a^{+})}{p(-a^{+})}=\frac{K_{Q}^{-}}{K_{P}^{-}} \; \text{,}
\end{equation}
\begin{equation}
\frac{q(a^{-})}{p(a^{-})}=\frac{K_{Q}^{+}}{K_{P}^{+}} \; \text{.}
\end{equation}

Thus, we intend to find the multipliers ($K_{P}^{-}, K_{P}^{+}, K_{Q}^{-}, K_{Q}^{+}$) from the finite element analysis output by plotting the ratios $\nicefrac{p(s)}{\sqrt{s}},\nicefrac{q(s)}{\sqrt{s}}$
and taking the limit $s\rightarrow0$, and similarly at the other end point. We can then find ($Q,\sigma$) from equations \eqref{trLHS} and \eqref{trRHS}, fulfilling step (iii). The other steps can be pursued as discussed previously.

\section{Problems when a moment is developed}\label{moment}

\hspace{0.4cm} We now turn to problems which involve the effects of asymmetry, which may stem from an externally applied moment or an inherent asymmetry of the contacting geometry. As we shall see, it is not possible to write down a closed-form expression for the traction ratio as we did for the symmetric case in Eq. (\ref{tractionratio}) and we will need to make use of an asymptotic description of the ratio near the contact edges similar to that outlined in \textsection\ref{asymptotes} in order to find the differential tension, $\sigma = \sigma_{\text{A}} - \sigma_{\text{B}}$. The other steps, however, translate directly into the asymmetric regime.

Suppose that we have a contact spanning {[}-$a\qquad b${]}, and we make a small change, $\Delta P$ in normal load together with a small
change in applied moment, $\Delta M$. The corresponding change in contact
pressure is given by \cite{Sackfield_2001}
\begin{equation}
\Delta p(x)=\frac{\Delta P}{\pi\sqrt{\left(a+x\right)\left(b-x\right)}}+\frac{4\Delta M\left(2x+a-b\right)}{\pi\left(a+b\right)^{2}\sqrt{\left(a+x\right)\left(b-x\right)}} \; \text{.}
\end{equation}

If, at the same time, there are small changes in shear force, $\Delta Q,$
and differential bulk tension, $\Delta\sigma,$ the change in shear
traction, $\Delta q(x),$ generated is given by 
\begin{equation}
\Delta q(x)=\frac{\Delta Q}{\pi\sqrt{\left(a+x\right)\left(b-x\right)}}+\frac{\Delta\sigma\left(2x+a-b\right)}{8\sqrt{\left(a+x\right)\left(b-x\right)}} \; \text{.}
\end{equation}

The contact law is now given by $a=a(P,M)$ and $b=b(P,M)$. Therefore
\begin{equation}
\mathrm{d}a=\frac{\partial a}{\partial P}\mathrm{d}P+\frac{\partial a}{\partial M}\mathrm{d}M\text{,}  \quad 
\mathrm{d}b=\frac{\partial b}{\partial P}\mathrm{d}P+\frac{\partial b}{\partial M}\mathrm{d}M\text{.}
\end{equation}

We demand a proportional load path (Eq. \eqref{proploading}), i.e. $\mathrm{d}M/M=\mathrm{d}P/P$, so
that 
\begin{equation}
\mathrm{d}a=\left[\frac{\partial a}{\partial P}+\frac{\partial a}{\partial M}\frac{\mathrm{d}M}{\mathrm{d}P}\right]\mathrm{d}P=\left[\frac{\partial a}{\partial P}+\frac{\partial a}{\partial M}\left(\frac{M}{P}\right)\right]\mathrm{d}P  \; \text{,}
\end{equation}
\begin{equation}
\mathrm{d}b=\left[\frac{\partial b}{\partial P}+\frac{\partial b}{\partial M}\frac{\mathrm{d}M}{\mathrm{d}P}\right]\mathrm{d}P=\left[\frac{\partial b}{\partial P}+\frac{\partial b}{\partial M}\left(\frac{M}{P}\right)\right]\mathrm{d}P  \; \text{,}
\end{equation}
and hence, in place of equation (\ref{eqn:PsymmIntegral}), we now have 
\begin{equation}
p(x)=\frac{1}{\pi}\int_{P_{x}}^{P}\left[\frac{1}{\sqrt{\left(a+x\right)\left(b-x\right)}}+\frac{4\left(2x+a-b\right)}{\left(a+b\right)^{2}\sqrt{\left(a+x\right)\left(b-x\right)}}\frac{\mathrm{d}M}{\mathrm{d}\tilde{P}}\right]\mathrm{d}\tilde{P} \label{eqn:PmomIntegral}  \; \text{,}
\end{equation}
and proportional loading also requires that $\mathrm{d}P/P=\mathrm{d}Q/Q=\mathrm{d}\sigma/\sigma$
giving
\begin{equation}
q(x)=\int_{P_{x}}^{P}\left[\frac{1}{\pi\sqrt{\left(a+x\right)\left(b-x\right)}}\frac{\mathrm{d}Q}{\mathrm{d}\tilde{P}}+\frac{\left(2x+a-b\right)}{8\sqrt{\left(a+x\right)\left(b-x\right)}}\frac{\mathrm{d}\sigma}{\mathrm{d}\tilde{P}}\right]\mathrm{d}\tilde{P}\label{eqn:QmomIntegral}  \; \text{.}
\end{equation}

Following the symmetric case from \textsection \ref{nomoment}, we would like to reveal the ratio 
\begin{equation}\label{tractionratioM}
\frac{q(x)}{p(x)}=\frac{\displaystyle{\int_{P_{x}}^{P}\left[\frac{1}{\sqrt{\left(a+x\right)\left(b-x\right)}}\frac{\mathrm{d}Q}{\mathrm{d}\tilde{P}}+\frac{\pi\left(2x+a-b\right)}{8\sqrt{\left(a+x\right)\left(b-x\right)}}\frac{\mathrm{d}\sigma}{\mathrm{d}\tilde{P}}\right]\mathrm{d}\tilde{P}}}{\displaystyle{\int_{P_{x}}^{P}\left[\frac{1}{\sqrt{\left(a+x\right)\left(b-x\right)}}+\frac{4\left(2x+a-b\right)}{\left(a+b\right)^{2}\sqrt{\left(a+x\right)\left(b-x\right)}}\frac{\mathrm{d}M}{\mathrm{d}\tilde{P}}\right]\mathrm{d}\tilde{P}}} \; \text{,}
\end{equation}
but it is difficult to proceed with the integrals on the right-hand side. In a real problem we will know both $p(x)$ and $q(x)$ together with 
\begin{equation}
\frac{Q}{P}= \,\lambda,  \quad \frac{M}{ aP}= \,\gamma \;\text{,} 
\end{equation}
from the force resultants Eq. (\ref{normal_equilibrium}, \ref{roational_equilibrium}, \ref{tangentialeq}), where $a$ has been chosen arbitrarily as means of normalisation. It is the quantity
\begin{align}
\frac{a\,\sigma}{P}=& \,\eta \; \text{,}
\end{align}
and hence $\sigma$, that we seek. Although we cannot evaluate Eq. \eqref{tractionratioM} for a general point in closed-form, we are able to find the limits of the traction ratio at each end of the contact, where $\eta$ is the only remaining unknown. The results, which are  generalisations of those found in \textsection \ref{asymptotes}, are 
\begin{equation}\label{asymptotesLHS}
\frac{q(-a^{+})}{p(-a^{+})}=\frac{1}{8}\frac{\left(1+\left(\frac{b}{a}\right)\right)^{2}\left[8\lambda-\pi\eta\left(1+\frac{b}{a}\right)\right]}{\left(1+\left(\frac{b}{a}\right)\right)^{2}-4\gamma\left(1+\frac{b}{a}\right)}  \; \text{,}
\end{equation}
\begin{equation}\label{asymptotesRHS}
\frac{q(b^{-})}{p(b^{-})}=\frac{1}{8}\frac{\left(1+\left(\frac{b}{a}\right)\right)^{2}\left[8\lambda+\pi\eta\left(1+\frac{b}{a}\right)\right]}{\left(1+\left(\frac{b}{a}\right)\right)^{2}+4\gamma\left(1+\frac{b}{a}\right)}  \; \text{,}
\end{equation}
where details of finding the limits can be found in the Appendix \ref{Appendix:asymptote}. We note that we actually only need one of Eq. (\ref{asymptotesLHS}) and Eq. (\ref{asymptotesRHS}) to determine $\eta$ and hence $\sigma$. In practice, however, the second equation serves as a excellent check on the accuracy of the results.
%The traction ratios are revealed by evaluating the finite element output near the contact edge so that the left-hand sides of Eq. \eqref{asymptotesLHS} and \eqref{asymptotesRHS} are obtained using the fitted asymptotic forms described in \textsection\ref{asymptotes}. 
%
%Turning to the question of the sum of the bulk stresses, numerically integrating Eqs. (\ref{Qstrain}, \ref{sigmastrain}) will reveal the surface direct strain contribution from the shear tractions,
%and hence the value of $\sigma_{\text{A}}+\sigma_{\text{B}}$ is obtained by evaluating Eq. \eqref{SigmaSigma}.

All that remains is to find the sum of the bulk stresses, $\sigma_{\text{A}}+\sigma_{\text{B}}$, which can be evaluated by considering the direct strain in the contact interface, as previously. The only small change is somewhat notational, since the presence of a moment means that we replace the term $\varepsilon_{xx}^{P}(x)$ in Eq. (\ref{PMstrain}) by $\varepsilon_{xx}^{P,M}(x)$.

Therefore, in summary, when there is asymmetry in the problem, the steps necessary to find the values of $P$, $M$, $Q$, $\sigma_{\text{A}}$ and $\sigma_{\text{B}}$ in the presence of a moment can be summarised as follows.
\begin{enumerate}[(I)]
	\item Find values for the contact tractions $p(x)$ and $q(x)$ from simulations or otherwise.
	\item The contact resultants for $P$, $M$, and $Q$ are found from integrating the tractions, using Eqs. (\ref{normal_equilibrium}, \ref{roational_equilibrium}, \ref{tangentialeq}).
	\item Obtain the near edge multipliers ($K_{P}^{-}, K_{P}^{+}, K_{Q}^{-}, K_{Q}^{+}$) from the finite element output by plotting the ratios $\nicefrac{p(s)}{\sqrt{s}},\nicefrac{q(s)}{\sqrt{s}}$ and taking the limit $s\rightarrow0$, and similarly at the other end point.
	\item Use the ratios of the asymptotic multipliers in order to obtain the left-hand sides of Eq. \eqref{asymptotesLHS} and \eqref{asymptotesRHS}, and, hence, a way of determining the value of $\eta$ and thus $\sigma = \sigma_{\text{A}} - \sigma_{\text{B}}$.
	\item Evaluate the individual contributions to the direct strain, $\varepsilon_{xx}$, using Eqs. (\ref{PMstrain}, \ref{Qstrain}, \ref{sigmastrain}) and subtract those from the numerical output for the total strain to determine the sum of the bulk stresses using Eq. \eqref{SigmaSigma}.
	\item Finally, output $\sigma_{\text{A}} = \nicefrac{\sigma}{2} + \nicefrac{\sigma_{\text{A}} + \sigma_{\text{B}}}{2}$ and $\sigma_{\text{B}} = \nicefrac{\sigma_{\text{A}} + \sigma_{\text{B}}}{2}- \nicefrac{\sigma}{2}$.
\end{enumerate}
This algorithm gives the values of the loads necessary to formulate an equivalent half-plane representation of the full contact problem (cf. Figure \ref{fig:half_plane_analysis}). We shall now present an example that illustrates the effectiveness of this methodology.

\subsection{Dovetail geometry example}\label{dovetail_exmaple}

\hspace{0.4cm}  We return to the example of a fan blade dovetail root contact displayed in Figure \ref{fig:half_plane_analysis} a), and examine the implementation of a numerical analysis in the complex scenario of this prototype problem. The joint is commonly used to connect fan blades to the rotor disk in jet engines. As the engine starts spinning, centrifugal and expansion forces, $F_{\text{C}}$ and $T$, will act on the blade and disk, respectively. Note that, here, we consider a symmetrical configuration for the sake of simplicity (i.e. $F_{V} = 0$), while in reality a vibrational component may act on the blade thereby inducing an asymmetry about the centre-axis of the blade and a \textit{steady state} variation of the local load components we are interested in.
 
\begin{figure}[b!]
	\centering
	\includegraphics[scale=0.34, trim= 0 0 0 0, clip]{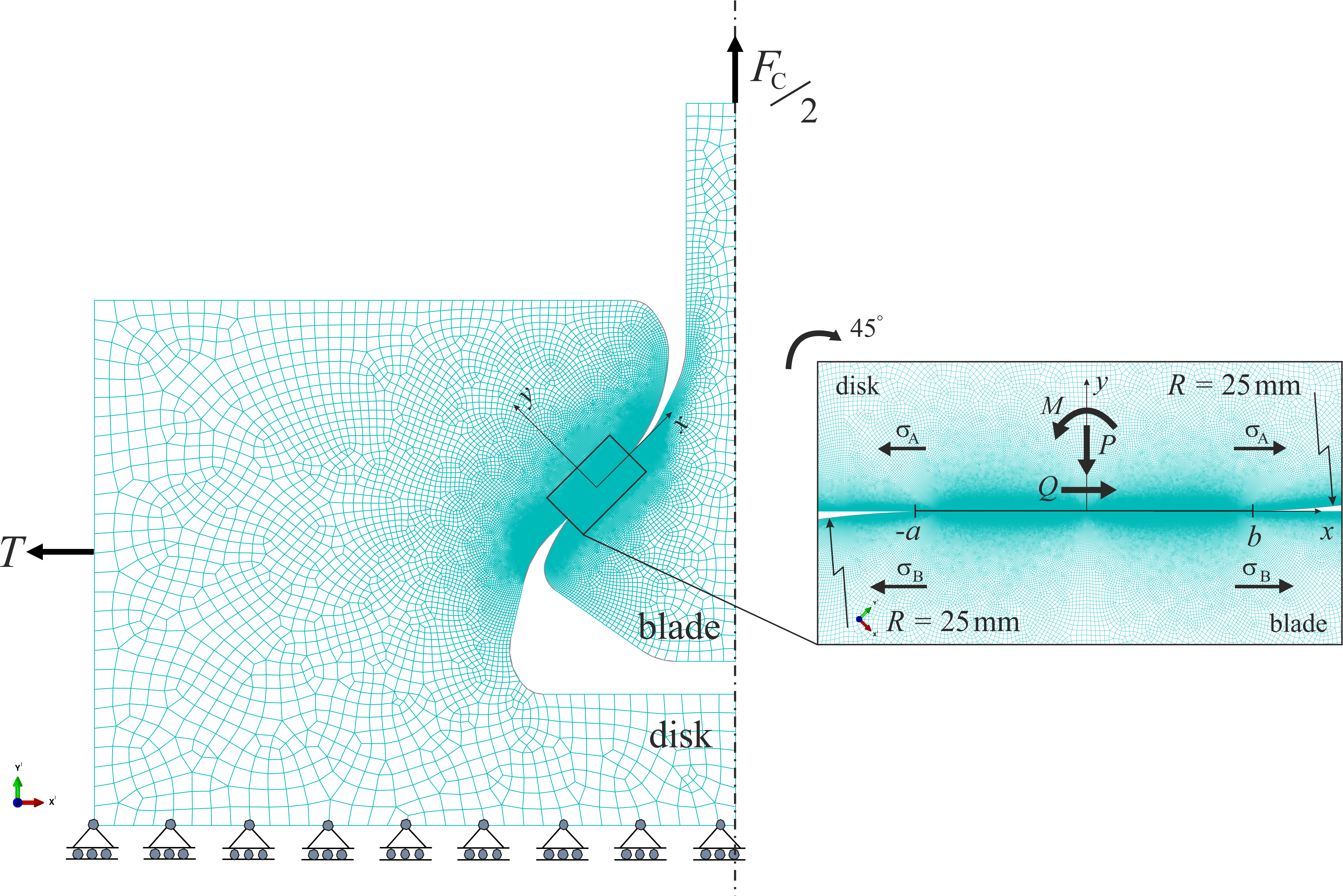}
	\caption{Finite element model of a dovetail prototype subject to external loads $F_{\text{C}}$ and $T$ and the local response.}
	\label{fig:dovetail_FE_Model}	
\end{figure}

The system, as depicted in Figure \ref{fig:dovetail_FE_Model}, is statically indeterminate. The equilibrium equations are insufficient to determine the local forces and reactions. As explained in the introduction, the finite element analysis of the structure allows us to connect the external loads to the local response where a matrix with complicated, yet unknown entries, makes this connection, see Eq. \eqref{matrix}. Our algorithm (I)-(VI) allows us to determine these local forces.

The finite element model consists of a disk and a blade-representation with symmetry conditions exploited about the vertical centre-axis and the disk supports forces in the vertical direction. As required by the method presented, the external loads, $F_\text{C}$ and $T$, are applied proportionally from an unloaded state under fully adhered conditions until they reach the desired load state. The structure is meshed with a gradual mesh refinement toward the contact interface and all other properties are the same as in the model used in \textsection\ref{Hertzexample}.

\begin{figure}[t!]
	\centering
	\includegraphics[scale=0.11, trim= 0 0 0 0, clip]{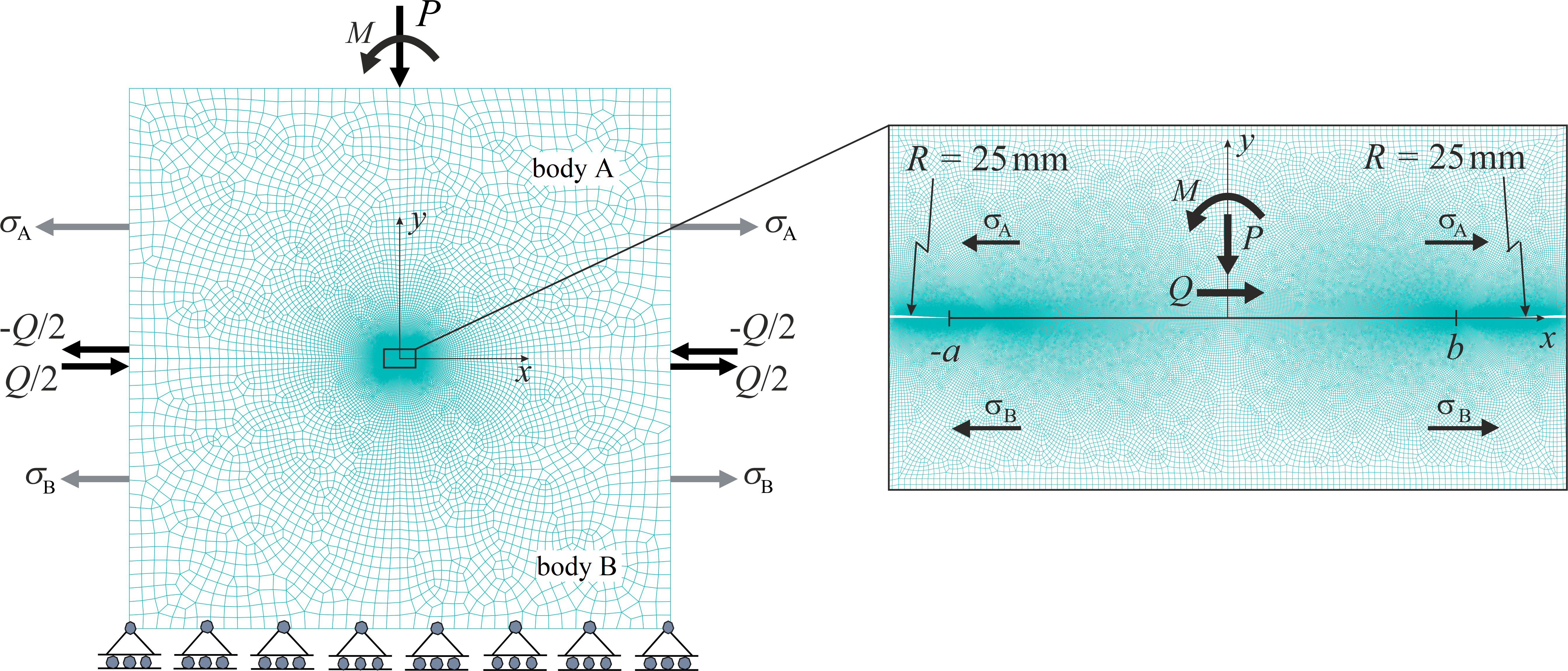}
	\caption{Finite element model of two half-planes connected by a flat and rounded geometry subject to loads $P$, $M$, $Q$, $\sigma_{\text{A}}$, and $\sigma_{\text{B}}$.}
	\label{fig:FRP_Fe_model}	
\end{figure}

For comparison, and in order to verify that the same contact behaviour is found in a half-plane representation of the dovetail root contact, we use the local loads found above to initiate an interface between two half-planes connected by a flat and rounded punch where the contacting shape is the same as the corresponding function for the dovetail root contact, see Figure \ref{fig:FRP_Fe_model}. In the half-plane representation, we expect to observe the same contact behaviour, up to a reasonable degree of accuracy, as in the prototype example. This means the normal and shear tractions, as well as the direct strain in the contact interface, ought to match between the two problems.

\begin{figure}[t!]
	\centering
	\includegraphics[scale=0.4, trim= 0 0 0 0, clip]{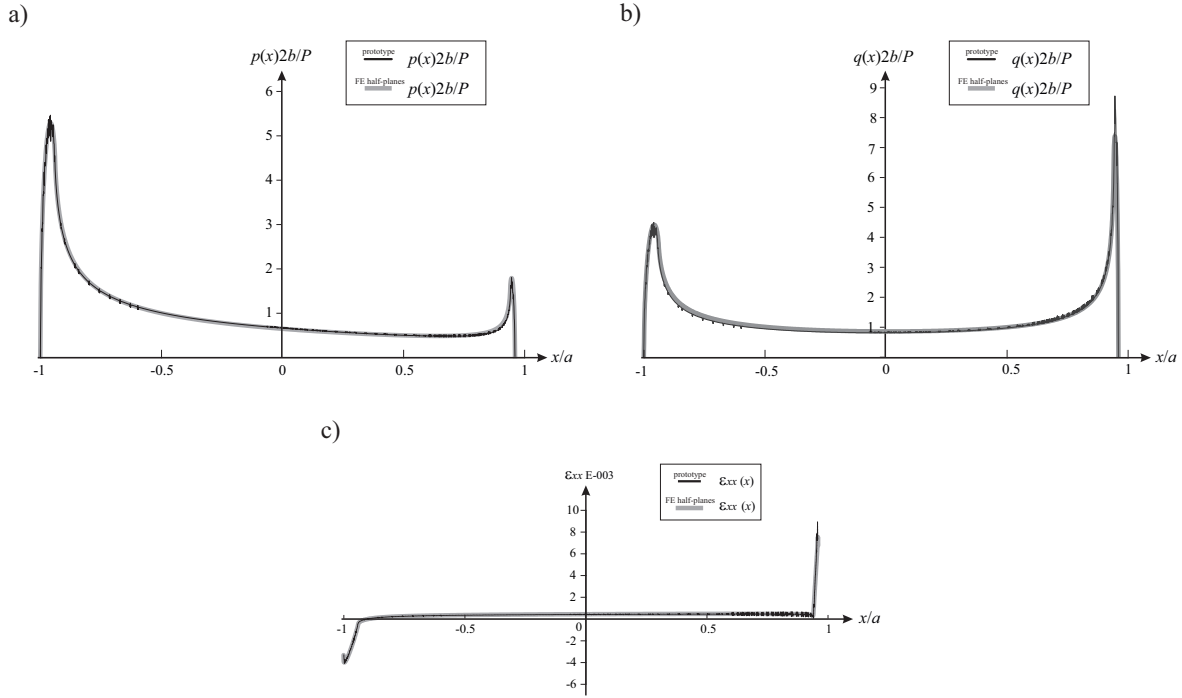}
	\caption{Comparison between finite element outputs for prototype, Figure \ref{fig:dovetail_FE_Model}, and the half-plane representation, Figure \ref{fig:FRP_Fe_model}, for a) the normal pressure distribution, $p(x)$, b) the shear traction distribution, $q(x)$, and c) the direct strain, $\varepsilon_{xx} (x)$, along the interface.}
	\label{fig:pressure_distribution}	
\end{figure}

The contact pressure and shear tractions calculated using the two finite element models are displayed in Figures \ref{fig:pressure_distribution} a) and b) as the black lines. It is trivial to use these to find the resultants $P$, $M$ and $Q$ using Eqs. \eqref{normal_equilibrium}--\eqref{tangentialeq}. To determine $\sigma_{\text{A}}$ and $\sigma_{\text{B}}$, since $M \neq 0$, we follow (III)-(VI) in using the edge asymptotes introduced in \textsection\ref{moment} to find the difference in bulk tensions, $\sigma$, and then use the direct strain, $\varepsilon_{xx}$, within the contact interface to reveal the sum of the bulk stresses, see Figure \ref{fig:pressure_distribution} c).

The prototype and half-plane representation differ from one another in terms of contact initialisation and mesh refinement. This may cause a slight difference in the quality of numerical results between the two. Notwithstanding, the normal tractions and contact extent shown in Figure \ref{fig:pressure_distribution} match up very well between prototype and half-plane. This indicates that the normal load, $P$, and moment, $M$, have been found accurately from the prototype.

The shear tractions are affected by the four load components, normal load, moment, shear load, and differential bulk tension. The comparison of shear tractions shown in Figure \ref{fig:pressure_distribution} b) and the direct strain in Figure \ref{fig:pressure_distribution} c) gives reassurance that the values derived for $Q$, $\sigma_{\text{A}}$ and $\sigma_{\text{B}}$ agree very well indeed between prototype and half-plane representation.  

\section{Conclusion}

\hspace{0.4cm} We have developed an algorithm which enables practical contact problems that are analysed, usually, by a finite element procedure using commercial code, to be matched to half-plane idealisations. We set ourselves the task of finding the best choices for the contact loads present in a prototype, that is the normal force $P$, shear force $Q$, moment $M$, and bulk stresses $\sigma_{\text{A}}, \sigma_{\text{B}}$. While exploiting equilibrium conditions to find the traction resultants $P$, $Q$ and $M$ is trivial, of particular note is the problem in devising the best-match for the bulk stresses acting
parallel with the surface of the contact, $\sigma_{\text{A}}$ and $\sigma_{\text{B}}$, whose difference is material in modifying the slip pattern, but which are separately responsible
for propelling cracks. We were able to extract these values by considering the traction ratio within the contact as well as the direct strains, which gave us two expressions for the difference and sum of the bulk stresses respectively. With the extracted values to hand, we are then able to formulate a half-plane problem to approximate the full prototype behaviour.

We considered two specific examples to illustrate the methodology. The first neglected all asymmetry (thus precluding a moment) and looked at the slightly academic set-up of a Hertzian prototype, in which the results are generally obtainable in closed form by other means. However, we were able to show that the algorithm was robust, finding excellent agreement with the expected values of $P$, $Q$, $\sigma_{\text{A}}$ and $\sigma_{\text{B}}$. The second example considered a prototype dovetail root geometry, in which there is a moment and formulating an equivalent half-plane problem is non-trivial. We used the algorithm to derive the contact loads, and were able to demonstrate that the methodology is again robust. 

Thus, given the results of commercial code, we are able to devise an equivalent problem using half-plane theory with suitable loads applied that can be used to analyse finer details of the contact problem that are poorly resolved by the  finite element analysis, for example the regions of slip at the contact edges. These results can then be used to set up a much simpler laboratory fretting fatigue test mimicking the prototype contact behaviour with the highest fidelity possible.

\section*{Acknowledgements}

\hspace{0.4cm}This project has received funding from the European
Union's Horizon 2020 research and innovation programme under the Marie
Sklodowska-Curie agreement No 721865. David Hills thanks Rolls-Royce
plc and the EPSRC for the support under the Prosperity Partnership
Grant 'Cornerstone: Mechanical Engineering Science to Enable Aero
Propulsion Futures', Grant Ref: EP/R004951/1. The authors would like to thank the comments of the anonymous referees that helped improve upon a previous version of this manuscript.

\bibliographystyle{elsarticle-num}
\bibliography{Contribs_References_Hendrik_Jan2019}

\begin{appendix}

%dummy comment inserted by tex2lyx to ensure that this paragraph is not empty

\section{Sequential loading}\label{S-sequentialloading}

Figure \ref{fig:PQSimga_Space} shows sequential and proportional loading for a $P, Q, \sigma$-problem neglecting the effect of a moment. The sequence in which the loads are applied greatly affects the procedure and the solution to a contact problem. For completeness, we look at the effects of a sequential load application in this appendix.

\begin{figure}[t!]
	\centering
	\includegraphics[scale=0.4, trim= 0 0 0 0, clip]{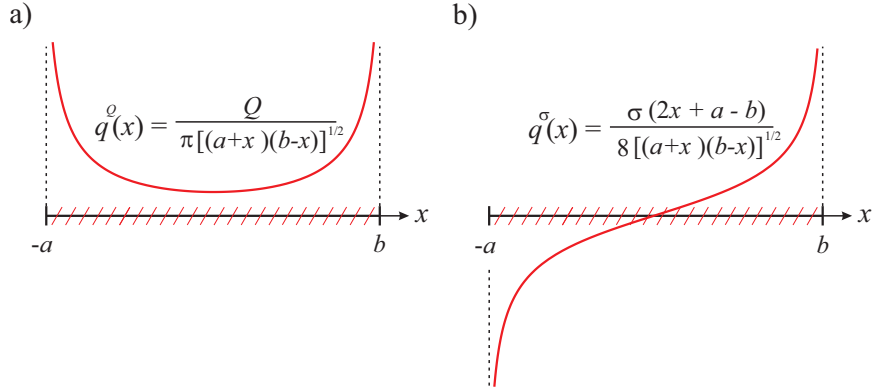}
	\caption{Fully adhered shear tractions due to a) a shear force, $Q$ and b) differential bulk tension, $\sigma$.}
	\label{fig:fully_adhered_cond}	
\end{figure}

Consider the problem shown in Figure \ref{fig:half_plane_analysis} b), and suppose
that the contacting bodies are elastically similar. We assume initially, that the coefficient of friction, $f$, is sufficiently high to inhibit all slip. 
%In a finite element analysis there is invariably the ability to achieve this by setting an appropriate switch. 
First, we apply the normal load, $P$ and potentially a moment, $M$, and establish a contact of extent [$-a\qquad b$].  In our analysis we will neglect the shear tractions induced by the application of the normal load as they will be much smaller in magnitude than the contact pressure when the bodies are elastically similar. Now we apply the external tangential loads ($Q,\sigma_{\text{A}},\sigma_{\text{B}}$). These will be square root singular in character at the contact edges - symmetric when induced by a shear force and antisymmetric when induced by differential bulk tension, as shown in Figure \ref{fig:fully_adhered_cond}. The full-stick shear traction distribution is given by
\begin{equation}\label{fullyadherededtractions}
q(x)=\frac{Q}{\pi\sqrt{\left(a+x\right)\left(b-x\right)}}+\frac{\sigma\left(2x+a-b\right)}{8\sqrt{\left(a+x\right)\left(b-x\right)}}\; \text{,}
\end{equation}
where $\sigma=\sigma_{\text{A}}-\sigma_{\text{B}}$. Suppose we introduce a new coordinate, $s=a+x$, measured from the left hand contact edge,
so that 
\begin{equation}
q(s)=\frac{Q}{\pi\sqrt{s\left(b+a-s\right)}}+\frac{\sigma\left(2s-a-b\right)}{8\sqrt{s\left(b+a-s\right)}} \; \text{,}
\end{equation}
and now take the limit $s\rightarrow0$, so that 
\begin{equation}\label{s}
q(s)\sqrt{s}\rightarrow\frac{Q}{\pi\sqrt{\left(a+b\right)}}-\frac{\sigma(a+b)}{8\sqrt{\left(a+b\right)}} \; \text{.}
\end{equation}

Similarly, if we measure $t=b-x$ from the left hand edge 
\begin{equation}
q(t)=\frac{Q}{\pi\sqrt{t\left(a+b-t\right)}}+\frac{\sigma(a+b-2t)}{8\sqrt{t\left(a+b-t\right)}} \; \text{,}
\end{equation}
and, in the limit $t\rightarrow0$ 
\begin{equation}\label{t}
q(t)\sqrt{t}\rightarrow\frac{Q}{\pi\sqrt{\left(a+b\right)}}+\frac{\sigma\left(a+b\right)}{8\sqrt{\left(a+b\right)}} \; \text{.}
\end{equation}

Therefore, by taking the output for $q(s)$ and $q(t)$ from the finite element analysis and
plotting the left hand sides of limits \eqref{s} and \eqref{t} we can deduce
the value of $Q$ and $\sigma$. Of course, we know the value of the
shear force, $Q$, by another route, equation \eqref{tangentialeq}, but this provides an independent
check to our results. 

We now turn our attention to a consideration of the absolute
bulk tensions present. The general expression for the strain induced
at the surface of a half-plane, under plane strain, by surface tractions, $p\left(x\right)$ and $q\left(x\right)$, is \cite{Barber_2010}
\begin{equation}\label{directstrain2}
\varepsilon_{xx}^{p(x), q(x)}(x)=-\frac{2}{\pi E^{*}}\int_{-a}^{b}\frac{q\left(\xi\right)\mathrm{d}\xi}{x-\xi}+\frac{\left(1-2\nu\right)}{(1-\nu)\,E^{*}}\,p(x)\; \text{,}
\end{equation}
where $E^{*}=E/\left(1-\nu^{2}\right)$, $E$ is Young's modulus
and $\nu$ is Poisson's ratio. We know the contact pressure rigorously, originating from the normal load, $P$, and moment, $M$. In any
practical problem, the total surface strains, $\varepsilon_{xx}$, everywhere on the surface can be found using the finite element output. The contribution from the contact pressure (and therefore $P, M$) is given by
\begin{equation}
\varepsilon_{xx}^{P,M}(x)=\frac{\left(1-2\nu\right)}{(1-\nu)\,E^{*}}\,p(x)\; \text{,}
\end{equation}
and as we have found a solution under no-slip conditions, so we can also
find the implied surface strains resulting from the effects of an applied shear force, $Q$, and the shear tractions arising from differential bulk tension, $\sigma$, rigorously, \textit{for once and for
	all}, from equation \eqref{fullyadherededtractions}, which gives 
\begin{equation}
\varepsilon_{xx}^{Q}(x)=-\frac{2Q}{\pi^{2}E^{*}}\int_{-a}^{b}\frac{\mathrm{d}\xi}{(x-\xi)\sqrt{\left(a+\xi\right)\left(b-\xi\right)}}=0\;, -a<x<b\; \text{,}
\end{equation}
and
\begin{equation}
\varepsilon_{xx}^{\sigma}(x)=-\frac{\sigma_{\text{A}}-\sigma_{\text{B}}}{4\pi E^{*}}\int_{-a}^{b}\frac{\left(2\xi+a-b\right)\mathrm{d}\xi}{\left(x-\xi\right)\sqrt{\left(a+\xi\right)\left(b-\xi\right)}}=-\frac{\sigma_{\text{A}}-\sigma_{\text{B}}}{4E^{*}}\;,-a<x<b\; \text{.}
\end{equation}

The total strain distribution within the fully stuck contact will
therefore have four contributions, \textit{viz}. the three just derived
$\varepsilon_{xx}^{P,M}(x),\varepsilon_{xx}^{Q}(x), \varepsilon_{xx}^{\sigma}(x)$
together with the effect of the \textit{average} remote bulk loads, i.e.
\begin{equation}\label{sigmasumstrain2}
\varepsilon_{xx}^{\sigma_{\text{A}}+\sigma_{\text{B}}}\left(x\right)=\frac{\sigma_{\text{A}}+\sigma_{\text{B}}}{2E^{*}}\; \text{,}
\end{equation}
giving, therefore, the sum of the bulk stresses. Notice that the results
derived here should be robust in the sense that they are averaged
and do not rely on `spot' values. With sequential loading they are not geometry dependent: under full stick conditions the shear traction distribution is fully defined
without knowing the relative profile of the contacting bodies.

\section{Leading-order asymptotic form of (\ref{eqn:PmomIntegral})--(\ref{eqn:QmomIntegral})} \label{Appendix:asymptote}

Consider the expression for the contact pressure for an indenter under
both an applied normal force, $P$, and an applied moment, $M$, as
given by (\ref{eqn:PmomIntegral}): 
\begin{equation}
p(x)=\frac{1}{\pi}\int_{P_{x}}^{P}\left[\frac{1}{\sqrt{\left(a+x\right)\left(b-x\right)}}+\frac{4\left(2x+a-b\right)}{\left(a+b\right)^{2}\sqrt{\left(a+x\right)\left(b-x\right)}}\frac{\mathrm{d}M}{\mathrm{d}\tilde{P}}\right]\mathrm{d}\tilde{P}.
\end{equation}
where $-a<x<b$. In this expression, the size of the contact
set is a function of both $P$ and $M$ with $a(P,M)=a$
and $b(P,M)=b$. We shall restrict ourselves to the case
in which the loading is proportional, and in particular, we set 
\begin{equation}
\frac{\mbox{d}M}{\mbox{d}P}=\gamma a.
\end{equation}
For notational brevity, we introduce the shorthand notation 
\begin{equation}
\bar{a}(P)=a(s,\gamma a s),\;\bar{b}(s)=b(s,\gamma a s).
\end{equation}

We seek the coefficients of the square-root bounded terms as $x\rightarrow-a^{+}$,
$x\rightarrow b^{-}$. We shall consider the right-hand contact
point in detail to illustrate the methodology, but a similar analysis
follows for the left-hand contact point. Suppose 
\begin{equation}
x=b-\delta X
\end{equation}
where $0<\delta\ll1$ and $X=O(1)$ strictly. Provided that $\bar{b}(P)$
is sufficiently well-behaved so as to have a Taylor
expansion there, we know that 
\begin{equation}
P_{x}=P-\delta\hat{P}\;\mbox{where}\;\hat{P}=\frac{X}{\bar{b}'(P)}+O(\delta).\label{eqn:hatP}
\end{equation}

In particular, since we shall only require the leading-order term
in the asymptotic expansion, it will be sufficient for us to assume
that $X-\hat{P}\bar{b}'(P)=o(1)$ henceforth. In the integral,
we set $\tilde{P}=P-\delta S$. Then, 
\begin{equation}
p=\frac{\sqrt{\delta}}{\pi}\int_{0}^{\hat{P}}\left(1+\frac{4\gamma a(b+a)}{(a+b)^{2}}+O(\delta)\right)\frac{1}{\sqrt{(a+b)+O(\delta)}}\frac{1}{\sqrt{X-S\bar{b}'(P)+O(\delta)}}\,\mbox{d}S.
\end{equation}

Assuming $\delta$ is small, we can expand each term in the integrand
in terms of Taylor series, finding that 
\begin{equation}
p=\frac{\sqrt{\delta}}{\pi\sqrt{a+b}}\left(1+\frac{4\gamma a}{(a+b)}\right)\int_{0}^{X/\bar{b}'(P)}\frac{1}{\sqrt{X-S\bar{b}'(P)}}\,\mbox{d}S+O(\delta^{3/2}),
\end{equation}
where we have used the approximation for $\hat{P}$ in (\ref{eqn:hatP}).
Thus, after integrating and returning to the original variables, we
see that 
\begin{equation}
p(x)=\frac{2}{\bar{b}'(P)\pi\sqrt{a+b}}\left(1+\frac{4\gamma a}{(a+b)}\right)\sqrt{b-x}+o\left(\sqrt{b-x}\right)
\end{equation}
as $x\rightarrow b^{-}$.

By setting $x=-a+\delta X$ where $0<\delta\ll1$ and $X$ is
strictly order unity, a similar analysis yields 
\begin{equation}
p(x)=\frac{2}{\bar{a}'(P)\pi\sqrt{a+b}}\left(1-\frac{4\gamma a}{(a+b)}\right)\sqrt{a+x}+o\left(\sqrt{a+x}\right)
\end{equation}
as $x\rightarrow-a^{+}$.

Naturally, a similar process also holds for the shear tractions, $q(x)$,
as given by (\ref{eqn:QmomIntegral}), 
\begin{equation}
q(x)=\int_{P_{x}}^{P}\left[\frac{1}{\pi\sqrt{\left(a+x\right)\left(b-x\right)}}\frac{\mbox{d}Q}{\mbox{d}\tilde{P}}+\frac{\left(2x+a-b\right)}{8\sqrt{\left(a+x\right)\left(b-x\right)}}\frac{d\Delta\sigma}{\mbox{d}\tilde{P}}\right]\mbox{d}\tilde{P}.
\end{equation}

We again assume proportional loading, defining 
\begin{equation}
\frac{\mbox{d}Q}{\mbox{d}P}=\lambda,\;\frac{\mbox{d}\Delta\sigma}{\mbox{d}P}=\frac{\eta}{a}.
\end{equation}

Then, an asymptotic analysis yields 
\begin{equation}
q(x)=\frac{2}{\bar{b}'(P)\sqrt{a+b}}\left(\frac{\lambda}{\pi}+\frac{\eta(a+b)}{8a}\right)\sqrt{b-x}+o\left(\sqrt{b-x}\right)
\end{equation}
as $x\rightarrow b^{-}$, and 
\begin{equation}
q(x)=\frac{2}{\bar{a}'(P)\sqrt{a+b}}\left(\frac{\lambda}{\pi}-\frac{\eta(a+b)}{8a}\right)\sqrt{a+x}+o\left(\sqrt{a+x}\right)
\end{equation}
as $x\rightarrow-a^{+}$.

Therefore, the desired ratios of the coefficients are given by 
\begin{equation}
\frac{q(-a^{+})}{p(-a^{+})}=\frac{1}{8}\frac{\left(1+\left(\frac{b}{a}\right)\right)^{2}\left[8\lambda-\pi\eta\left(1+\frac{b}{a}\right)\right]}{\left(1+\left(\frac{b}{a}\right)\right)^{2}-4\gamma\left(1+\frac{b}{a}\right)},
\end{equation}
as we approach the left-hand contact point, and 
\begin{equation}
\frac{q(b^{-})}{p(b^{-})}=\frac{1}{8}\frac{\left(1+\left(\frac{b}{a}\right)\right)^{2}\left[8\lambda+\pi\eta\left(1+\frac{b}{a}\right)\right]}{\left(1+\left(\frac{b}{a}\right)\right)^{2}+4\gamma\left(1+\frac{b}{a}\right)},
\end{equation}
as we approach the right-hand contact point, respectively. 
\end{appendix}
\end{document}